\documentclass[a4paper,11pt]{article}

\usepackage{pos}
\usepackage{lipsum} 
\usepackage[outercaption]{sidecap}   
\usepackage{gensymb}
\usepackage{xcolor,lineno}

\modulolinenumbers[5]

\title{The Hybrid Elevated Radio Observatory for Neutrinos (HERON) Project}

\ShortTitle{GRAND-BEACON aka HERON}

\author*[a,b]{Kumiko Kotera}
\author[]{Ingo Allekotte}
\author[]{Jaime Alvarez-Muñiz}
\author[]{Sergio Cabana-Freire}
\author[]{Valentin Decoene}
\author[]{Luciano Ferreyro}
\author[]{Arsène Ferrière}
\author[]{Matias Hampel}
\author[]{Olivier Martineau-Huynh}
\author[]{Valentin Niess}
\author[]{Federico Sanchez}
\author[]{Stephanie Wissel}
\author[]{Andrew Zeolla}
\onbehalf{for the BEACON and GRAND Collaborations{\normalsize \\ \normalfont(a complete list of affiliations and authors can be found at the end of the proceedings)}\\}

\affiliation[a]{Sorbonne Universit\'{e} et CNRS, UMR 7095, Institut d'Astrophysique de Paris, 98 bis bd Arago, 75014 Paris, France}
\affiliation[b]{Vrije Universiteit Brussel (VUB), Dienst ELEM, Pleinlaan 2, B-1050, Brussels, Belgium}

\emailAdd{kotera@iap.fr}

\abstract{

Measuring ultra-high energy neutrinos, with energies above $10^{16}$\,eV, is the next frontier of the emerging multi-messenger era. Their detection requires building a large-scale detector with 10 times the instantaneous sensitivity of current instruments, sub-degree angular resolution, and wide daily field of view. The Hybrid Elevated Radio Observatory for Neutrinos (HERON) is designed to be that discovery instrument. HERON combines the complementary features of two radio techniques being demonstrated by the BEACON and GRAND prototypes. Its preliminary design consists of 24 compact, elevated phased stations with 24 antennas each, embedded in a sparse array of 360 standalone antennas. This setup tunes the energy threshold to below 100\,PeV, where the neutrino flux should be high. The sensitivity of the phased stations combines with the powerful reconstruction capacities of the standalone antennas to produce an optimal detector. HERON is planned to be installed at an elevation of 1,000\,m across a 72 km-long mountain range overlooking a valley in Argentina’s San Juan province. It would be connected to the worldwide network of multimessenger observatories and search for neutrino bursts from candidate sources of cosmic rays, like gamma-ray bursts and other powerful transients. With HERON’s deep sensitivity, this strategy targets discoveries that cast new light into the inner workings of the most violent astrophysical sources at uncharted energies. We present the preliminary design, performances, and observation strategy of HERON.

\vspace{4mm}

}

\ConferenceLogo{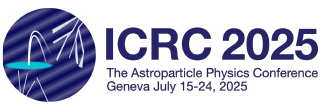}

\FullConference{39th International Cosmic Ray Conference (ICRC2025)\\  15--24 July, 2025\\ Geneva, Switzerland}

\begin{document}

\maketitle

Astrophysics and particle physics predict a low flux of UHE neutrinos necessitating gigantic observatories for detection. The Pierre Auger Observatory has established through the parent UHE cosmic ray measurements that, if the models used to interpret UHECR observations are adequate, then the flux of cosmogenic neutrinos is likely low~\cite{PierreAuger:2019azx}. This indicates that astrophysical neutrinos may be a more promising target, requiring a new observation strategy. IceCube has successfully detected hundreds of high-energy ($\lesssim$\,PeV) neutrinos over the past decade. However, only 2 sources could be identified over the isotropic background with 10 years of statistics~\cite{IC_Science_2022, IceCube_Science_2023}, due to the limited (few degree) angular resolution. Large-scale UHE neutrino projects for radio in-ice, in air, optical, and direct particle detection 
are planned, but over the coming decades.  Among them, two experiments in prototyping stages, BEACON~\cite{Wissel_2020} and GRAND~\cite{Martineau:GRAND_ICRC25,  batista2024giantradioarrayneutrino}, target the impulsive radio signals in the 10-1000\,MHz range emitted by particle cascades in the atmosphere from neutrino interactions in the ground. These radio signals, emitted in a narrow relativistic cone, can travel hundreds of kilometers with minimal atmospheric attenuation. Radio antenna-based instruments are ideal for achieving large detection areas due to their scalability, robustness, and cost-effectiveness, as demonstrated by the successful detection of UHE cosmic-ray air-showers in the AugerPrime Radio array.

HERON plans to simultaneously:
{a)} have enough sensitivity to detect UHE neutrinos, 
{b)} achieve excellent angular resolution to pinpoint sources amid a near-isotropic background, 
{c)} target the 100\,PeV neutrino energy range to extend and connect with IceCube observations \cite{2013PhRvL.111b1103A}, 
and, {d)} to be integrated within a multi-messenger framework, enabling rapid response to follow-up bursting sources while also sending alerts~\cite{2019ARNPS..69..477M}. 

Building on existing prototypes, HERON aims at demonstrating a large-scale, fully autonomous radio antenna array, achieving low detection thresholds and sub-degree angular resolution, rejecting radio interference, and distinguishing neutrinos from UHE cosmic ray background. It would be complementary to in-ice radio experiments, IceCube-Gen2 Radio and its precursor RNO-G, which will have wider fields of view but shallower instantaneous sensitivities.

\section{A hybrid design for enhanced detection}\label{sec1}

\begin{SCfigure}
\includegraphics[width=0.6\textwidth]{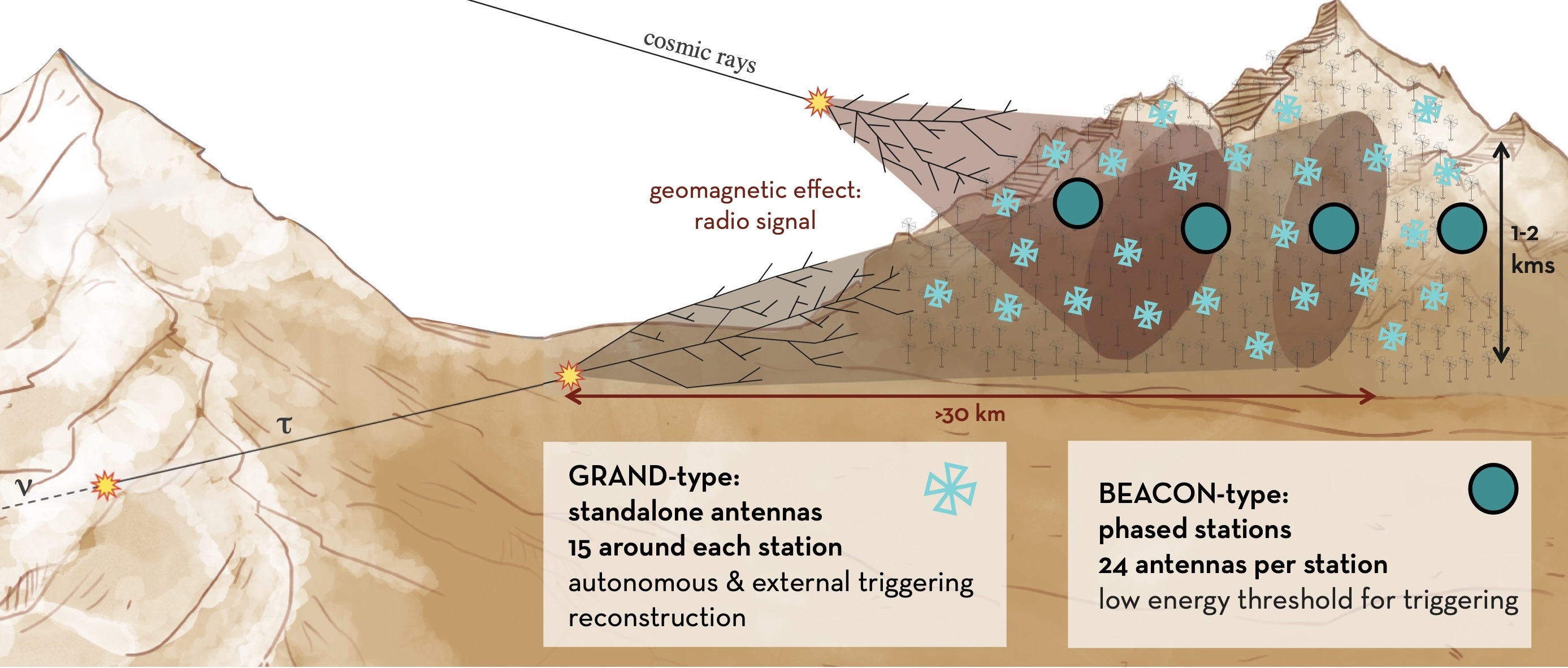}
\caption{\footnotesize UHE tau neutrinos ($\nu$) produced at such sources interact underground and create a tau lepton particle ($\tau$) that exits into the atmosphere and decays. The ensuing air shower emits a radio signal detected by antennas. HERON will combine phased dense antenna arrays to trigger on weaker signals, and sparse arrays of standalone antennas for reconstruction.}
\label{tab:instr_char}
\end{SCfigure}

The HERON concept combines the strengths of BEACON and GRAND radio techniques into a unique hybrid detector (Fig.\,\ref{tab:instr_char}). The BEACON-inspired phased antenna array detects weaker signals, and its high elevation increases the observable area for neutrino interactions. However, higher altitudes limit detection to more energetic particle showers due to the distance. A 1000\,m altitude balances these factors. The GRAND-like sparse array complements this by improving high-energy sensitivity, aiding particle reconstruction, and enhancing background noise rejection. Together, these features create a cost-effective instrument for identifying and reconstructing neutrino-induced events emerging from just below the Earth's surface. The details of the optimization of these parameters by full simulations is described in \cite{HERON_sims_ICRC25}.

\begin{SCfigure}
\includegraphics[width=0.49\textwidth,trim={0.5cm 0.5cm 0.3cm 0},clip]{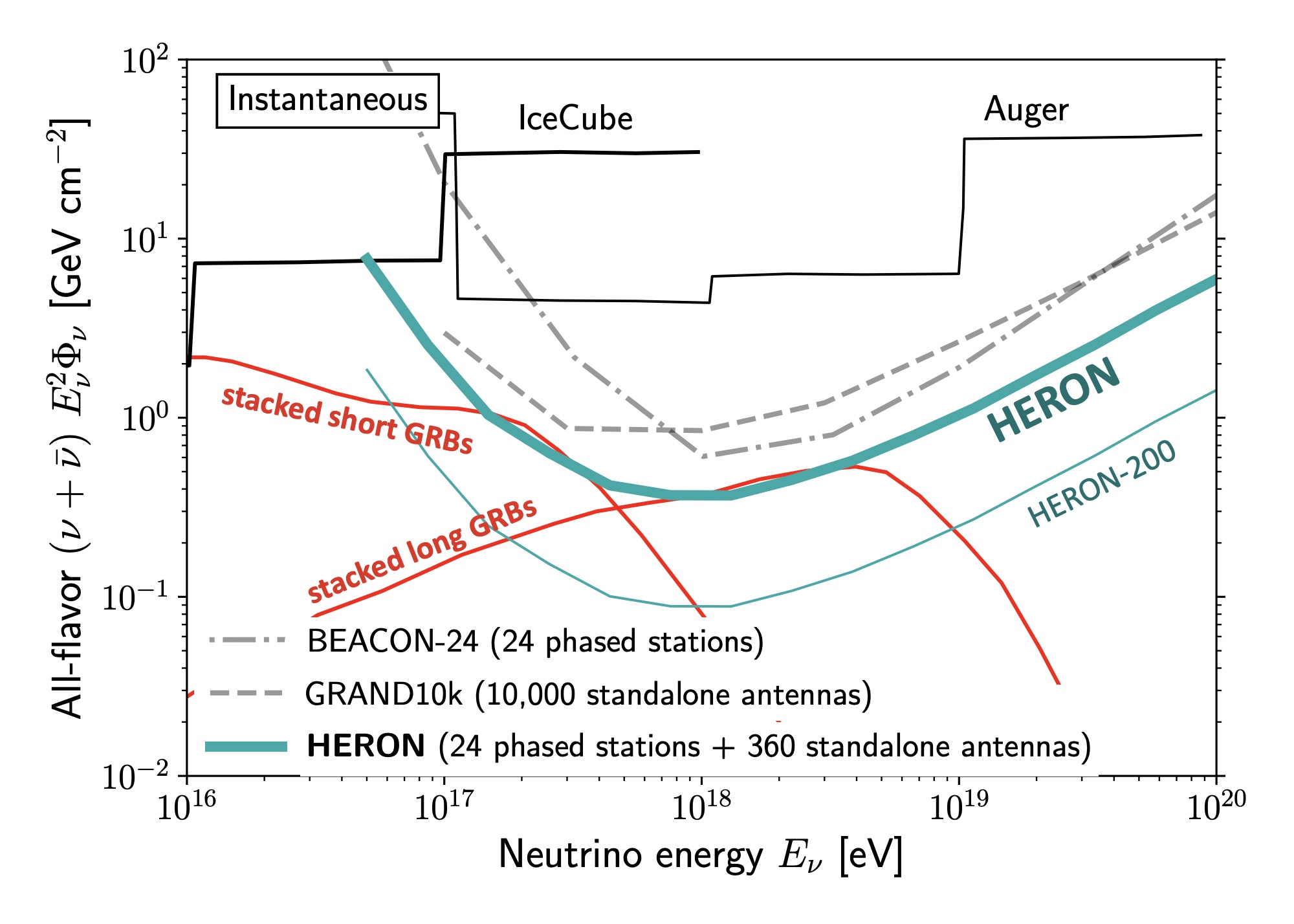}
 \vspace{-0.7cm}
 \caption{\label{fig:heron_sens} \footnotesize The expected instantaneous fluence sensitivities of HERON, GRAND,  BEACON, and a scaled-up HERON-200 for the future (see text) are illustrated alongside the predicted fluences from a stacked population of 200 short and long gamma-ray bursts (GRBs). With fewer than 1,000 antennas in total and 24 phased stations, HERON will operate at lower energies where the neutrino flux is more abundant, achieving sensitivity comparable to the mid-scale projects GRAND10k (10,000 antennas), BEACON-24 (24 phased stations), or to the large-scale project IceCube-Gen2 Radio. HERON will be a discovery instrument.}
\end{SCfigure}

In its preliminary design (Fig.~\ref{tab:instr_char}), HERON features an array of 24 BEACON-like individually \textit{phased stations} separated by 3 km on a mountain slope at an altitude above ground of 1000 m over a distance of 72 km, overlooking a valley in the San Juan province in Argentina. Each station is equipped with 24 high-gain dual-polarized antennas. The phased antennas will digitally delay and sum signals to enhance sensitivity towards the horizon. Between each of the BEACON-like stations, 15 \textit{standalone antennas} will be deployed at altitudes ranging from 500 m to 1500 m. These 360 standalone antennas can operate autonomously, as in the GRAND concept, and can also receive triggers from the BEACON-like stations. They are also designed for high gain with a beam focused toward the horizon, where neutrino-induced air showers are anticipated.  With fewer than 1,000 antennas, this innovative setup will achieve greater sensitivity to neutrinos than GRAND10k (10,000 antennas) and BEACON-24 (240 antennas across 24 stations) as shown in Fig.\,\ref{fig:heron_sens}. HERON's sensitivity also extends to lower energies, below 100 PeV, where the neutrino flux is more abundant, targeting compelling astrophysical models of neutrino production, Fig.\,\ref{fig:heron_sens}.

\subsection{Phased array design} 

The HERON phased arrays are designed as an evolution of the existing BEACON prototype. This setup includes 24 active dipoles placed within a small area ($100-200$\,m between the longest baseline). These dipoles are connected by coaxial cables to a data acquisition (DAQ) system, which forms beams to generate the trigger. A noise-riding trigger will continuously monitor noise levels, dynamically adjusting the trigger parameters to maintain a consistent rate of 10 Hz. Upon triggering, the system captures full waveforms from the phased antennas at a sampling rate of 500 MS/s and transmits a trigger to the 30 nearby standalone antennas via long-range WiFi.
Each dipole antenna consists of T-bars  
providing a gain of $25-30$\,dB and inline frequency filtering. 
The frequency range, initially set between $30-80$\,MHz is under study for performance optimization. It depends on the signal-to-noise (SNR) ratio and the practicality of different antenna designs \cite{wissel2024targeting100pevtauneutrino,Huege_2024}. 
The DAQ system includes digitizer boards with  analog-to-digital converters (ADCs) and an onboard FPGA to manage the low threshold trigger (Fig.~\ref{fig:blockdiag}). 
Digitizer/trigger boards independently form beams for both horizontal and vertical polarizations. 
A controller board connects to the digitizer and trigger boards through a single-board computer, managing power and communication systems and transmits the external trigger to the nearby standalone antennas.

\subsection{Standalone antennas design}
For the hardware (antenna, electronics, communications, power supply) and DAQ (Fig.~\ref{fig:blockdiag}) HERON's standalone antenna design, the reference would be the GRANDProto300 detection units observing in the $50-200$\,MHz frequency range, given the successful results from the GRANDProto300 commissioning phase \cite{GP300_ICRC25}. Enhancements are being explored, such as the potential advantages of utilizing the lower 30–80 MHz frequency range, where most air shower radio emissions—and sky background—are concentrated. This approach could enable the use of lower-cost ADCs and SoCs, reducing power consumption. Conversely, it may lower the peak signal\,\cite{AlvarezMuniz:2012sa}, impacting air shower identification methods. Additionally,  antenna designs are investigated, that enhance gain towards the horizon while minimizing it towards the sky, thereby improving the detector's SNR for neutrino-induced air shower \cite{wissel2024targeting100pevtauneutrino,Huege_2024}. 
Tilting the antenna pole is also under consideration to maximize its directional sensitivity towards phased array beams, adapting dynamically in response to external alerts from transient sources.

\subsection{Multiple trigger systems} Phased array triggers would be formed independently both for vertical 
and horizontal polarisation by computing a power sum for each beam direction as is done in the existing BEACON prototype.  
The first level phased trigger requires an impulsive signal 
in the beam in a time-window consistent with a bandlimited pulse. A second level trigger
would require both that the signal is purely linearly polarized and present in both vertical and horizontal polarisations as expected from radio emission in air showers in Argentina given the orientation of the antennas, expected shower directions and magnetic field. Multiple beams would be formed pointing within a band centered at the horizon. 
A feedback process would be implemented that monitors the local continuous noise in the channels and raises the thresholds 
automatically if there is a loud source of anthropogenic background.

In our preliminary design under optimization, phased-array triggers would be transmitted to the 45 adjacent standalone antennas through direct communication for the 15 closest units, and through the central DAQ and then the two neighboring phased arrays for the 30 surrounding antennas. 
The digitized data from the standalone antennas will be stored in a sufficiently long buffer (typically a few milliseconds) to enable the phased array trigger to be generated and received. Subsequently, a time trace centered around the anticipated signal arrival will be transmitted to the 
closest phased array DAQ.
The phased arrays and nearby standalone antennas would be synchronized to the ns level accuracy by phase-aligning sine waves broadcasted at different frequencies by three beacons~\cite{PierreAuger:2015aqe}. 
This both allows the individually triggered standalone antennas to synchronize to the phased arrays, and the internal boards of the phased array to synchronize together and share a central clock.
Standalone antennas will also self-trigger based on the GRAND concept. Advanced algorithms (NUTRIG \cite{Correa:NUTRIG_ICRC25} and denoising \cite{BenoitLevy:denoising_ICRC25}) may enable detection of low SNR signals, potentially improving sensitivity beyond the limits shown in Fig.\,\ref{fig:skymap}.

\vspace{-0.3cm}
\subsection{San Juan: an adequate deployment site for radio background-noise and topography} 
\begin{figure}
\centering
    \vspace{-0.7cm}
\includegraphics[width=0.49\textwidth]{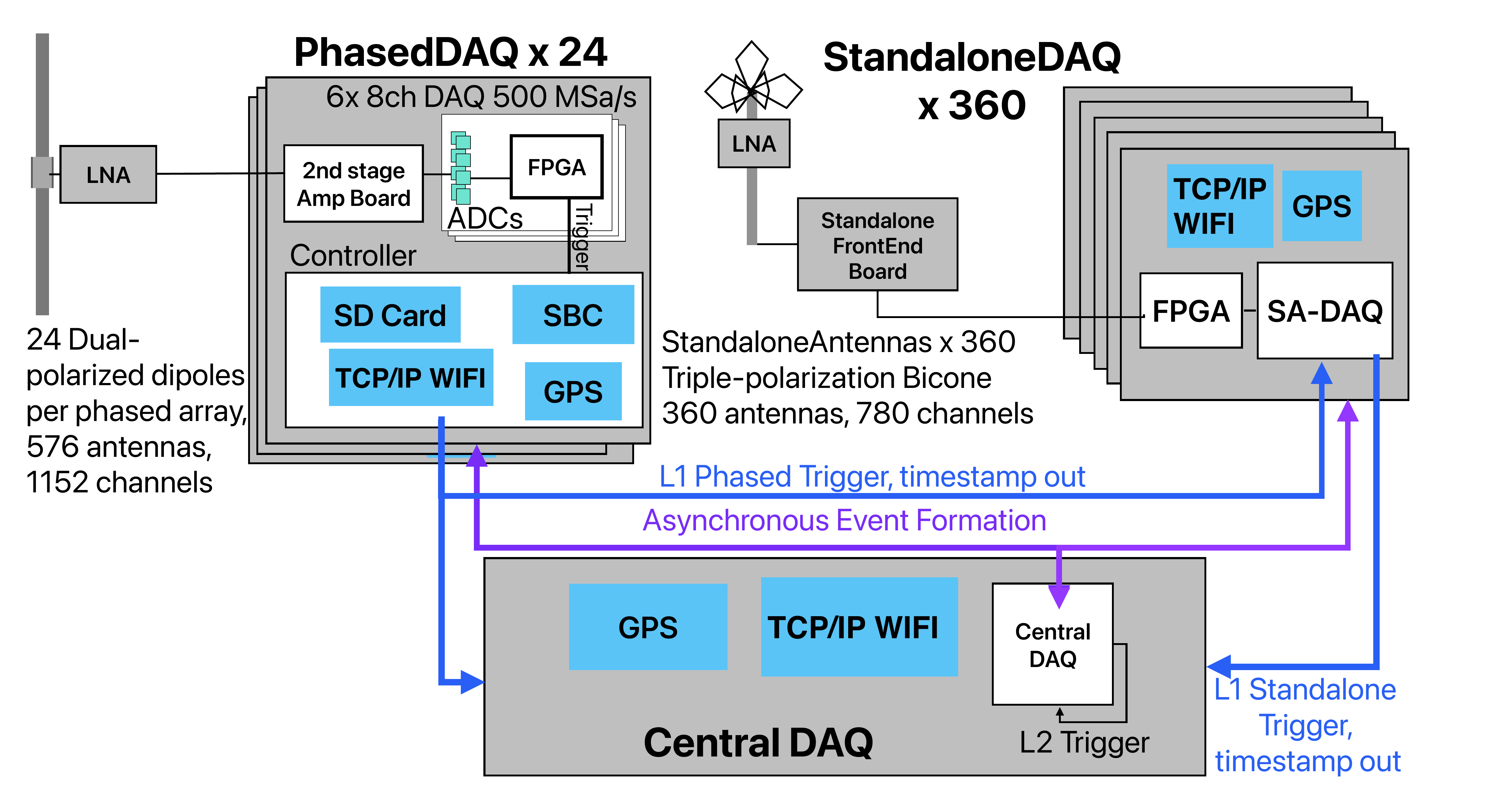}
\includegraphics[width=0.5\textwidth,trim={0.5cm 0 0.7cm 0.7cm},clip]{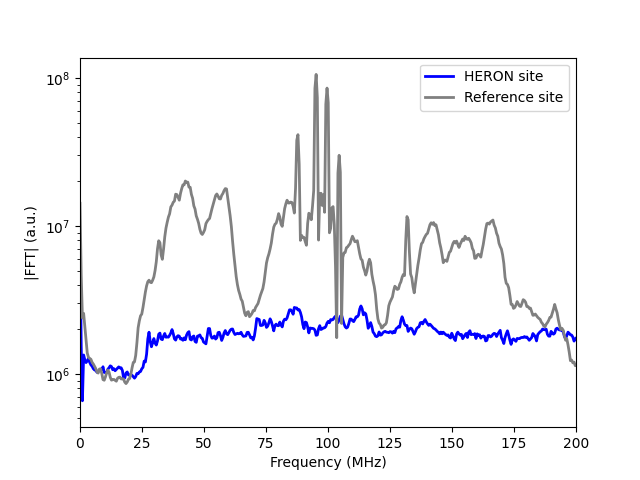}
    \caption{{\bf Left:} Block diagram of the HERON electronics and multiple triggering system. {\bf Right:} Power spectrum distributions (in arbitrary units, a.u.) measured at the San Juan site (blue) and at a reference site closer to the San Juan city (grey) with the same setup.}
    \label{fig:blockdiag}
\vspace{-0.3cm}

\end{figure}

\begin{figure}
    \centering
    \vspace{-0.2cm}
    \includegraphics[height=0.38\linewidth]{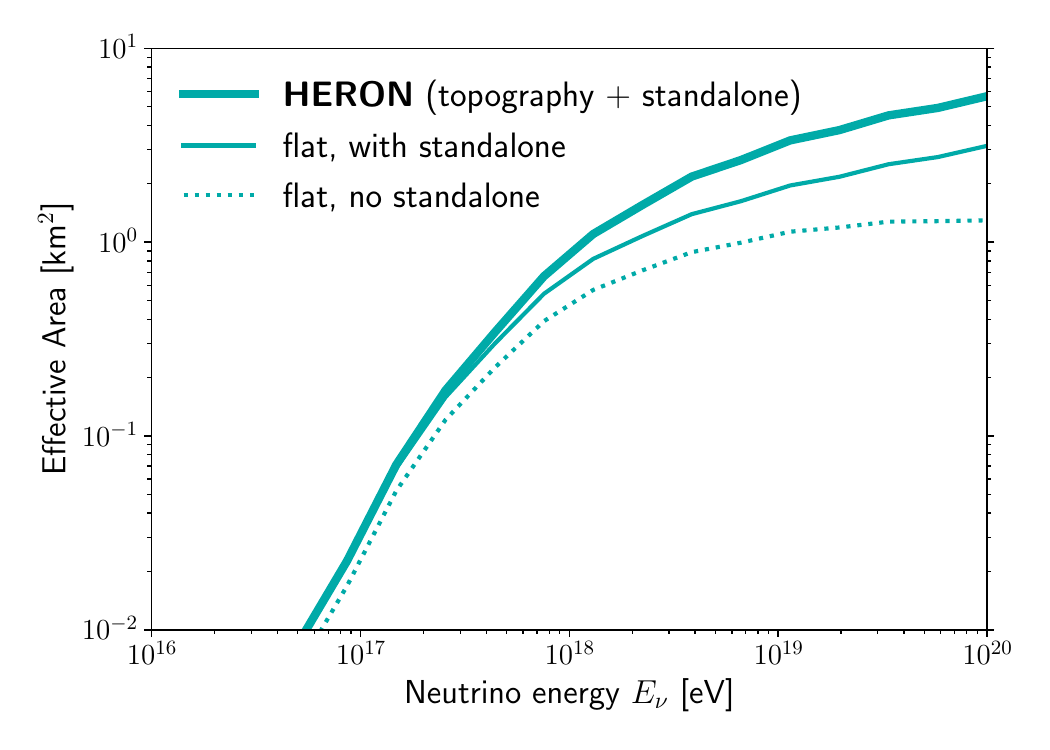}
    \raisebox{0.5cm}{\includegraphics[height=0.33\linewidth]{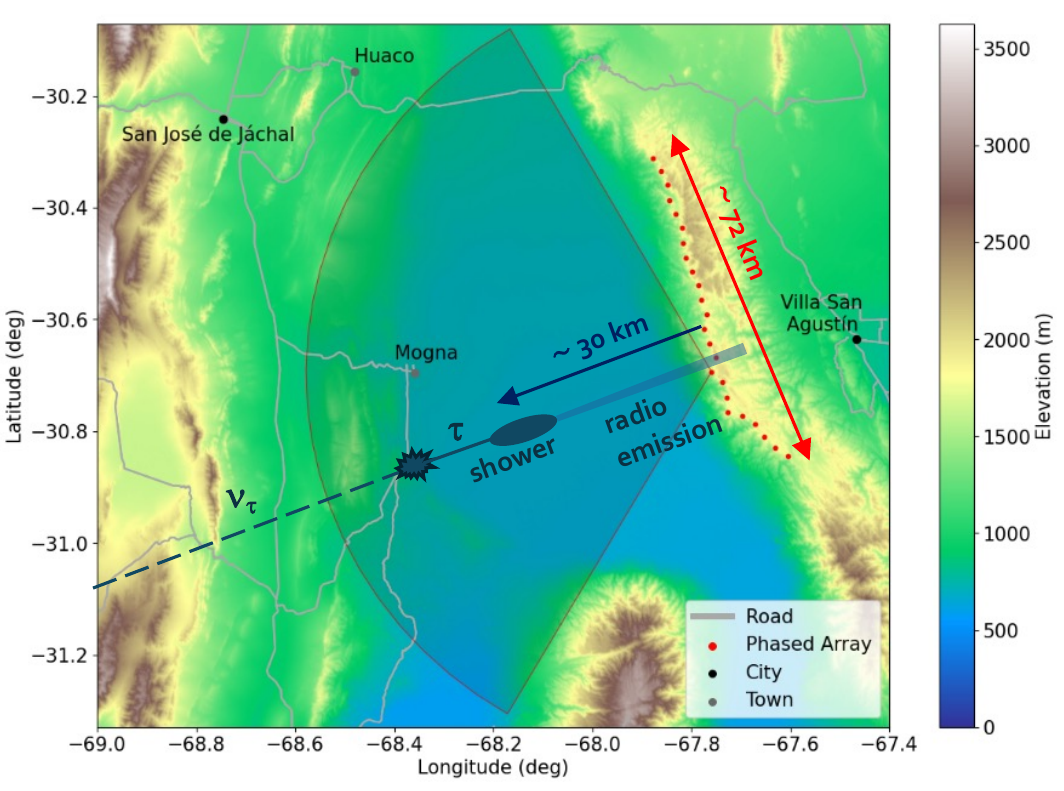}}
   \vspace{-0.3cm}
    \caption{
    {\bf Left:} Standalone antennas enable a factor $\,\gtrsim 2$ gain at UHE on the HERON effective area. The San Juan topography (right panel) enhances effective area by another factor of $\sim 2$ compared to a dense-rock spherical Earth case. 
    {\bf Right:} Site map of the San Juan area. Terrain is colored by elevation. Red dots represent the location of 24 phased arrays on the Eastern ridge, facing West overlooking a valley. A 100 km, 120 degree cone illustrates the field-of-view for radio signal of each station. 
    }
    \label{fig:site}
    \vspace{-0.6cm}
\end{figure}

An adequate site in the San Juan province in Argentina has been identified~(Fig.\,\ref{fig:site} right).  It presents the required conditions in terms of RFI background noise as shown in the frequency spectra taken on site (Fig.\,\ref{fig:blockdiag}, right) during a preliminary survey in April 2022. The topography is also ideal: a valley running North-South of width 30-60\,km, spanning 120\,km (we need 70\,km). 
Both ridges culminate at height $\sim 2000\,$m with a valley altitude $\sim 800\,$m. This encased topography enables to enhance the detection of neutrinos, thanks to the elevation and the larger target for neutrino interaction.

\vspace{-0.3cm}
\section{Simulated performances}
\vspace{-0.3cm}

\subsection{Offline neutrino reconstruction and identification} 
Reconstruction of neutrino properties would mostly rely on the standalone antenna data, that provide a wider footprint sample. A selection procedure for neutrino identification is being developed based on key features: the air shower trajectory of neutrino-induced air shower should point below the horizon, the polarisation of the signal should be consistent with that expected from the geomagnetic effect and the arrival direction\,\cite{Chiche_polar}, and the showers maximum of development should be deeper in atmosphere than for cosmic rays. 

{\it For self-triggered standalone data}, one can build on features such as the size of the Cherenkov ring  or the radius of curvature of the radio wavefront\,\cite{Decoene:2021atj}. For angular reconstruction, performances better than $0.1^{\circ}$ are achievable for spaced arrays according to simulations\,\cite{guelfand_2025}.

{\it For data triggered with the phased array}, one can use beamforming methods~\cite{Schoorlemmer:2020low, PierreAuger:2023opi}. 
Interferometric reconstruction is standard for the phased array antennas~\cite{Southall:2022yil, Romero-Wolf:2014pua}, in part because, by design the antennas view the same portion of the shower. 

\begin{figure}
    \centering
    \vspace{-0.7cm}
\includegraphics[width=0.99\linewidth]{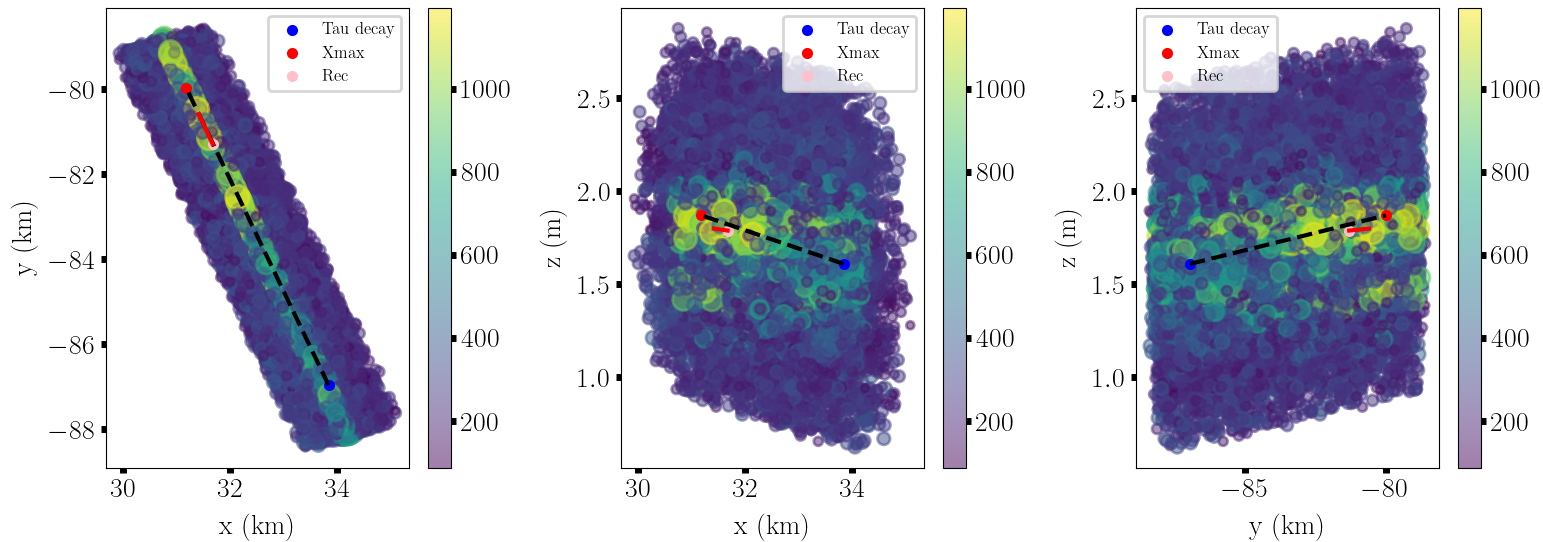}
    \caption{ Example of an event reconstruction using the beamforming scan for the standalone array. Each dot represents a location from where the phasing is tested the color and scale match the maximum of the beamformed signal achieved for that specific source location. The three different panels are projections of this scanning along the X-Y plane ({\it left}), X-Z plane ({\it middle}), and Y-Z plane ({\it right}). The dotted black line linking the tau decay position to the $X_{\rm max}$ location displays the shower direction, and the solid red line represents the reconstructed direction. From this beamforming map, we clearly see the direction of propagation of the shower and its region of maximal development, where the coherence signal is the strongest.}
    \label{fig:beamformed_rec}
    \vspace{-0.4cm}
\end{figure}

We preliminarily assessed a beamforming sparse array in HERON using simulations of $\sim 1,000$ tau-decay events with energy $2\times10^7-10^9$\,GeV induced by neutrinos~\cite{Decoene:2019izl,HERON_sims_ICRC25} with DANTON\,\cite{Niess:2018opy}.
Simulations were run over a realistic topography (of the San Juan valley) and array layout of 65 antennas. Electric fields were computed in frequency ranges of 50-200\,MHz, accounting for a galactic noise of $\sigma=22\,\mu$V/m (since ground noise is negligible in this frequency range). The standalone antennas are beamformed assuming coherent point source-like emission from the EAS. The signals from each antenna are phased accordingly to a spherical wavefront from a specific location. In order to reconstruct the direction of the EAS we proceed to a beamforming scan, where several locations are tested for phasing and the maximum of the beamformed signal is saved. This results in a 3D mapping of the maximal coherence of the signal across the space surrounding the array, as displayed in Fig.~\ref{fig:beamformed_rec}. From these mappings, we extract the direction of the EAS, with an accuracy of $0.4\degree$ on average. Note that these simulations are run for the standalone antenna array only, and the results improve when data from the phased array are added. (More details in \cite{HERON_sims_ICRC25}.)

\subsection{HERON's capabilities as a transient neutrino observatory}
\begin{figure}
 \centering 
 \vspace{-0.7cm}
 \includegraphics[height=0.33\textwidth]{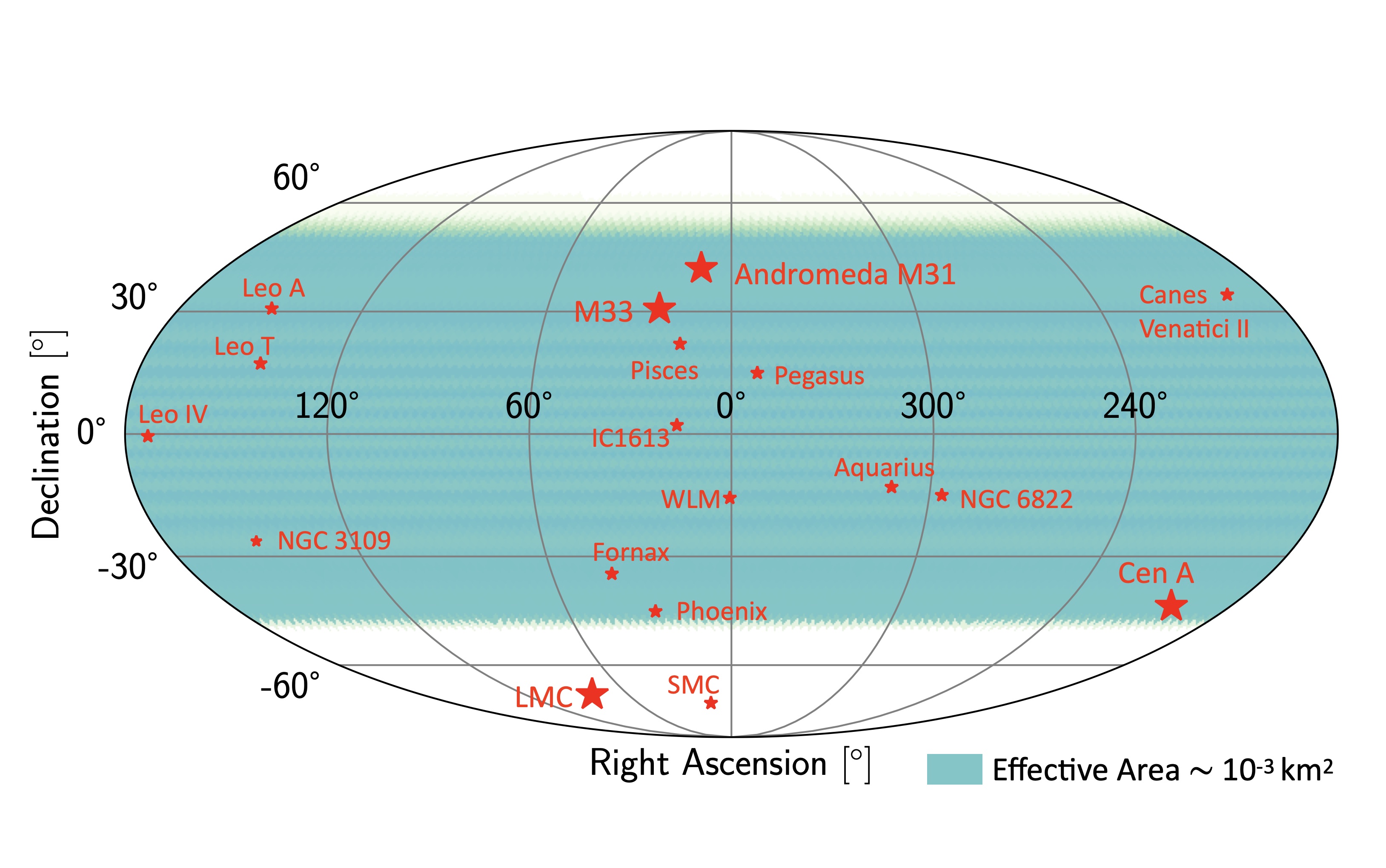}
 \includegraphics[height=0.33\textwidth]{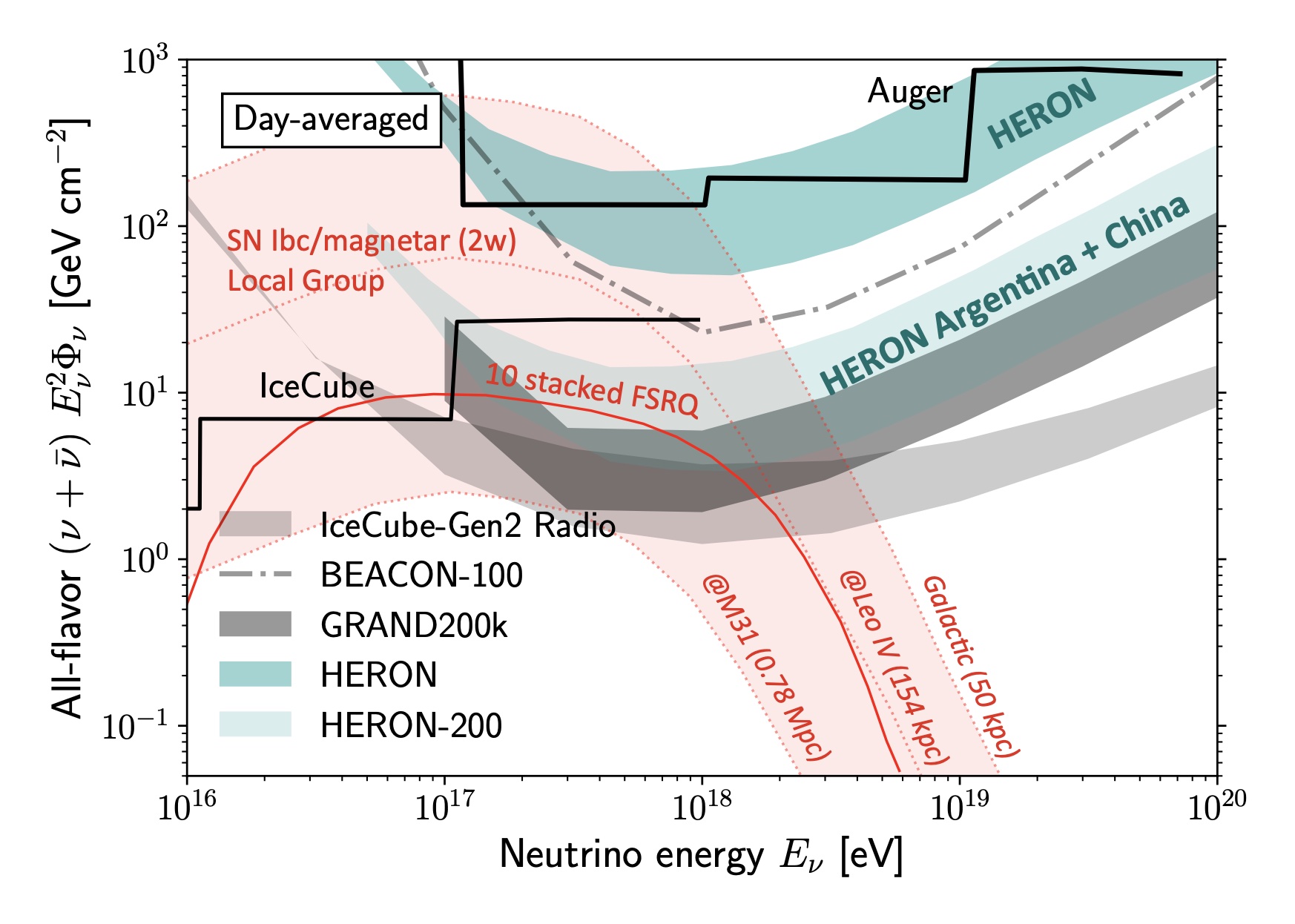}
 \vspace{-0.5cm}
\caption{\label{fig:skymap}\footnotesize HERON will have the potential to observe UHE neutrinos from transient sources occurring in our Local Group. It would be sensitive to magnetars or SNIbc occurring out to $\sim 200\,$kpc. {\rm \bf Left:} Daily averaged sky coverage of HERON. 
Overlayed are the positions of the nearest galaxies in our Local Group, that could harbor SNIbc and of the nearby active galactic nucleus Cen A (a promising source of UHE neutrinos \cite{Murase_2024}). Most are in the HERON field of view. {\rm \bf Right:}  HERON fluence sensitivity (blue band) to long bursts (> 30 min), spanning declinations of $-45^\circ$ to $+55^\circ$. 
For illustration, we show the maximal sensitivity for a scaled up 200-phased-station HERON assuming 4 sites (two mountain ridges facing in different directions with 50 stations each), and the sensitivity for GRAND200k (200,000 antennas in a single site). In red and pink: the expected neutrino fluences (adapted from \cite{Murase:2009pg}) for a SNIbc accompanied by a young magnetar 2 weeks (2w) after the explosion, located at various distances in the Local Group (at 50\,kpc in the Galaxy, in the Leo IV galaxy, or in the Andromeda galaxy, M31). Red solid line: neutrino fluence predicted for 10 stacked flares of Flat Spectrum Radio Quasars (FSRQs) at redshift $z = 2$~\cite{oikonomou2021multimessenger}.
} 
\vspace{-0.4cm}
\end{figure}

As illustrated in Figs.~\ref{fig:heron_sens}  and \ref{fig:skymap}, HERON could be an optimal instrument for discovering transient neutrino emissions through an astrophysics-informed approach, thanks to its {\it deep} instantaneous sensitivity (probing high-redshift), albeit with a narrow field of view. 
By combining astrophysical and multi-messenger data, HERON would focus on likely neutrino sources, enabling the detection of powerful and rare transients such as short and long GRBs, binary neutron star mergers, and young magnetars.
For instance, HERON could discover in a $5-10$ year observation time neutrinos of energy $\sim$\,300 (3000) PeV  from a realistic sample of 200 short-duration (long) GRBs linked to binary neutron star mergers (GRB afterglows)  \cite{2022NatRP...4..697G, Kimura:2017kan,Murase:2007yt}, a number extrapolated  from the rate of GRBs triggered by Fermi GBM and SVOM gamma-ray satellites, folded in HERON's 6\%-sky instantaneous field of view. Population studies show that nearby events would have longer durations \cite{2019ARNPS..69..477M,2022NatRP...4..697G}, with a high probability of also being observed by HERON thanks to its 70\%-sky daily field of view. With its expected sub-degree angular resolution and multi-messenger framework, HERON could conduct target-of-opportunity searches and effectively manage alerts from other observatories.

To highlight the potential of the instrument, we also show the sensitivity of HERON-200, which corresponds to a scaled up version of HERON with 200 stations. An option for station distribution is: over 2 facing mountain ridges instrumented with 50 antennas each, in Argentina and in China. The instantaneous field of view scales with the number of sites $N_{\rm site}$, hence the operation time required to collect 200 triggered GRBs scales down as $N_{\rm site}$. 

\vspace{-0.3cm}

\section{Conclusion}\label{sec3}
\vspace{-0.3cm}
\noindent The combined approach of phased triggering arrays and autonomous and sparse reconstruction antennas has several advantages. The phased arrays allow the experiment to target lower energies than in either the BEACON or GRAND designs. The beams can be turned directly on the horizon and directional masking of individual beams can be used for background rejection. The sparse antennas are spread out across longer baselines. As such, they can enhance event reconstruction, discrimination between neutrinos and cosmic rays and other backgrounds. They also provide additional effective area at high energies. Future work will focus on further optimizing the hardware design, the layout and the local topography, and quantifying the array reconstruction capability.

\bibliographystyle{ICRC}
\setlength{\bibsep}{0pt plus 0.3ex}
{\footnotesize
\bibliography{references}

\providecommand{\href}[2]{#2}\begingroup\raggedright\begin{thebibliography}{10}

\bibitem{PierreAuger:2019azx}
{\bfseries Pierre Auger} Collaboration, A.~Aab {\em et~al.} \href{http://dx.doi.org/10.1088/1475-7516/2019/11/004}{{\em JCAP} {\bfseries 11} (2019) 004}.

\bibitem{IC_Science_2022}
R.~Abbasi, M.~Ackermann, J.~Adams, {\em et~al.} \href{http://dx.doi.org/10.1126/science.abg3395}{{\em Science} {\bfseries 378} no.~6619, (Nov., 2022) 538–543}.

\bibitem{IceCube_Science_2023}
R.~Abbasi, M.~Ackermann, J.~Adams, {\em et~al.} \href{http://dx.doi.org/10.1126/science.adc9818}{{\em Science} {\bfseries 380} no.~6652, (June, 2023) 1338–1343}.

\bibitem{Wissel_2020}
S.~Wissel {\em et~al.} \href{http://dx.doi.org/10.1088/1475-7516/2020/11/065}{{\em JCAP} {\bfseries 11} (2020) 065}.

\bibitem{Martineau:GRAND_ICRC25}
{\bfseries GRAND} Collaboration, O.~Martineau {\em PoS} {\bfseries ICRC2025} (2025) .

\bibitem{batista2024giantradioarrayneutrino}
{GRAND Collaboration}, ``{The Giant Radio Array for Neutrino Detection (GRAND) Collaboration -- Contributions to the 10th International Workshop on Acoustic and Radio EeV Neutrino Detection Activities (ARENA 2024)},'' in {\em {10th International Workshop on Acoustic and Radio EeV Neutrino Detection Activities}}.
\newblock 2024.
\newblock \href{http://arxiv.org/abs/2409.03427}{{\ttfamily arXiv:2409.03427 [astro-ph.IM]}}.

\bibitem{2013PhRvL.111b1103A}
M.~G. {Aartsen}, R.~{Abbasi}, Y.~{Abdou}, M.~{Ackermann}, J.~{Adams}, J.~A. {Aguilar}, M.~{Ahlers}, D.~{Altmann}, J.~{Auffenberg}, X.~{Bai}, and et~al. \href{http://dx.doi.org/10.1103/PhysRevLett.111.021103}{{\em Physical Review Letters} {\bfseries 111} no.~2, (July, 2013) 021103}.

\bibitem{2019ARNPS..69..477M}
K.~{Murase} and I.~{Bartos} \href{http://dx.doi.org/10.1146/annurev-nucl-101918-023510}{{\em Annual Review of Nuclear and Particle Science} {\bfseries 69} (Oct., 2019) 477--506}.

\bibitem{HERON_sims_ICRC25}
{\bfseries BEACON and GRAND} Collaboration, A.~Zeolla {\em PoS} {\bfseries ICRC2025} (2025) .

\bibitem{wissel2024targeting100pevtauneutrino}
S.~Wissel, A.~Zeolla, C.~Deaconu, V.~Decoene, K.~Hughes, Z.~Martin, K.~Mulrey, A.~Cummings, R.~A. Batista, A.~Benoit-Lévy, M.~Bustamante, P.~Correa, A.~Ferrière, M.~Guelfand, T.~Huege, K.~Kotera, O.~Martineau, K.~Murase, V.~Niess, J.~Zhang, O.~Krömer, K.~Plant, and F.~G. Schroeder, ``Targeting 100-pev tau neutrino detection with an array of phased and high-gain reconstruction antennas,'' 2024.
\newblock \url{https://arxiv.org/abs/2409.02042}.

\bibitem{Huege_2024}
T.~Huege and O.~Krömer \href{http://dx.doi.org/10.1088/1748-0221/19/11/P11022}{{\em Journal of Instrumentation} {\bfseries 19} no.~11, (Nov, 2024) P11022}.

\bibitem{GP300_ICRC25}
{\bfseries GRAND} Collaboration, Pengxiong, Y.~Zhang, P.~Zhang, and X.~Xu {\em PoS} {\bfseries ICRC2025} (2025) .

\bibitem{AlvarezMuniz:2012sa}
J.~\'Alvarez-Mu\~niz, W.~R. Carvalho, A.~Romero-Wolf, M.~Tueros, and E.~Zas \href{http://dx.doi.org/10.1103/PhysRevD.86.123007}{{\em Phys.\ Rev.\ D} {\bfseries 86} (2012) 123007}.

\bibitem{PierreAuger:2015aqe}
{\bfseries Pierre Auger} Collaboration, A.~Aab {\em et~al.} \href{http://dx.doi.org/10.1088/1748-0221/11/01/P01018}{{\em JINST} {\bfseries 11} no.~01, (2016) P01018}.

\bibitem{Correa:NUTRIG_ICRC25}
{\bfseries GRAND} Collaboration, P.~Correa {\em PoS} {\bfseries ICRC2025} (2025) .

\bibitem{BenoitLevy:denoising_ICRC25}
{\bfseries GRAND} Collaboration, A.~Benoit-Lévy {\em PoS} {\bfseries ICRC2025} (2025) .

\bibitem{Chiche_polar}
S.~Chiche, K.~Kotera, O.~Martineau-Huynh, M.~Tueros, and K.~D. de~Vries \href{http://dx.doi.org/10.1016/j.astropartphys.2022.102696}{{\em Astropart. Phys.} {\bfseries 139} (2022) 102696}.

\bibitem{Decoene:2021atj}
V.~Decoene, O.~Martineau-Huynh, and M.~Tueros \href{http://dx.doi.org/10.1016/j.astropartphys.2022.102779}{{\em Astropart. Phys.} {\bfseries 145} (2023) 102779}.

\bibitem{guelfand_2025}
M.~Guelfand, V.~Decoene, O.~Martineau-Huynh, S.~Prunet, M.~Tueros, O.~Macias, and A.~Benoit-Lévy \href{http://dx.doi.org/https://doi.org/10.1016/j.astropartphys.2025.103120}{{\em Astroparticle Physics} {\bfseries 171} (2025) 103120}.

\bibitem{Schoorlemmer:2020low}
H.~Schoorlemmer and W.~R. Carvalho \href{http://dx.doi.org/10.1140/epjc/s10052-021-09925-9}{{\em Eur. Phys. J. C} {\bfseries 81} no.~12, (2021) 1120}.

\bibitem{PierreAuger:2023opi}
{\bfseries Pierre Auger} Collaboration, A.~Abdul~Halim {\em et~al.} \href{http://dx.doi.org/10.22323/1.444.0380}{{\em PoS} {\bfseries ICRC2023} (2023) 380}.

\bibitem{Southall:2022yil}
D.~Southall {\em et~al.} \href{http://dx.doi.org/10.1016/j.nima.2022.167889}{{\em Nucl. Instrum. Meth. A} {\bfseries 1048} (2023) 167889}.

\bibitem{Romero-Wolf:2014pua}
A.~Romero-Wolf {\em et~al.} \href{http://dx.doi.org/10.1016/j.astropartphys.2014.06.006}{{\em Astropart. Phys.} {\bfseries 60} (2015) 72--85}.

\bibitem{Decoene:2019izl}
V.~Decoene, N.~Renault-Tinacci, O.~Martineau-Huynh, D.~Charrier, K.~Kotera, S.~Le~Coz, V.~Niess, M.~Tueros, and A.~Zilles \href{http://dx.doi.org/10.1016/j.nima.2020.164803}{{\em Nucl. Instrum. Meth. A} {\bfseries 986} (2021) 164803}.

\bibitem{Niess:2018opy}
V.~Niess and O.~Martineau-Huynh.

\bibitem{2022NatRP...4..697G}
C.~{Gu{\'e}pin}, K.~{Kotera}, and F.~{Oikonomou} \href{http://dx.doi.org/10.1038/s42254-022-00504-9}{{\em Nature Reviews Physics} {\bfseries 4} no.~11, (Nov., 2022) 697--712}.

\bibitem{Kimura:2017kan}
S.~S. Kimura, K.~Murase, P.~M{\'e}sz{\'a}ros, and K.~Kiuchi \href{http://dx.doi.org/10.3847/2041-8213/aa8d14}{{\em Astrophys.\ J.} {\bfseries 848} no.~1, (2017) L4}.

\bibitem{Murase:2007yt}
K.~Murase \href{http://dx.doi.org/10.1103/PhysRevD.76.123001}{{\em Phys.\ Rev.\ D} {\bfseries 76} (2007) 123001}.

\bibitem{Murase_2024}
R.~Mbarek, D.~Caprioli, and K.~Murase.

\bibitem{Murase:2009pg}
K.~Murase, P.~Meszaros, and B.~Zhang \href{http://dx.doi.org/10.1103/PhysRevD.79.103001}{{\em Phys.\ Rev.\ D} {\bfseries 79} (2009) 103001}.

\bibitem{oikonomou2021multimessenger}
F.~Oikonomou, M.~Petropoulou, K.~Murase, A.~Tohuvavohu, G.~Vasilopoulos, S.~Buson, and M.~Santander \href{http://dx.doi.org/10.1088/1475-7516/2021/10/082}{{\em JCAP} {\bfseries 10} (2021) 082}.

\end{thebibliography}\endgroup
}
%

\clearpage

\section*{Full Author List: BEACON Collaboration (July 1st, 2025)}

\scriptsize
\noindent
J.~Alvarez-Mu\~{n}iz$^{1}$,
S.~Cabana-Freire$^{1}$,
W.~Carvalho~Jr.$^{2}$,
A.~Cummings$^{3,4,5}$,
C.~Deaconu$^{6}$,
J.~Hinkel$^{3}$,
K.~Hughes$^{7}$,
R.~Krebs$^{3,4}$,
Y.~Liu$^{7}$,
Z.~Martin$^{8}$,
K.~Mulrey$^{9,10}$
A.~Nozdrina$^{7}$,
E.~Oberla$^{6}$,
S.~Prohira$^{11}$,
A.~Romero-Wolf$^{12}$,
A.~G.~Vieregg$^{6,8,13}$,
S.~A.~Wissel$^{3,4,5}$,
E.~Zas$^{1}$,
A.~Zeolla$^{3,4}$
\\
\\
\noindent
$^{1}$Instituto Galego de F\'\i sica de Altas Enerx\'\i as IGFAE, Univerisade de Santiago de Compostela, 15782 Santiago de Compostela, Spain \\
$^{2}$Faculty of Physics, University of Warsaw, 02-093, Warsaw, Poland \\
$^{3}$Department of Physics, Pennsylvania State University, University Park, PA 16802, USA \\
$^{4}$Center for Multimessenger Astrophysics, Institute of Gravitation and the Cosmos, Pennsylvania State University, University Park, PA 16802, USA \\
$^{5}$Department of Astronomy and Astrophysics, Pennsylvania State University, University Park, PA 16802, USA \\
$^{6}$Department of Astronomy and Astrophysics, Kavli Institute for Cosmological Physics, University of Chicago, Chicago, IL 60637, USA \\
$^{7}$,Department of Physics, The Ohio State University, Columbus, OH 43210, USA \\
$^{8}$Department of Physics, Kavli Institute for Cosmological Physics, University of Chicago, Chicago, IL 60637, USA \\
$^{9}$Department of Astrophysics / IMAPP, Radboud University Nijmegen, 6500 GL, Nijmegen, The Netherlands \\
$^{10}$NIKHEF, Science Park Amsterdam, 1098 XG, Amsterdam, The Netherlands \\
$^{11}$Department of Physics and Astronomy, University of Kansas, Lawrence, KS 66045, USA \\
$^{12}$Jet Propulsion Laboratory, California Institute for Technology, Pasadena, CA 91109, USA \\
$^{13}$Enrico Fermi Institute, University of Chicago, Chicago, IL 60637, USA

\subsection*{Acknowledgments}
\noindent
This work is supported by NSF Awards $\#$ 2033500, 1752922,
1607555, and DGE-1746045 as well as the Sloan Foundation,
the RSCA, the Bill and Linda Frost Fund at the California Polytechnic State University, and NASA (support through JPL and Caltech as well as Award $\#$ 80NSSC18K0231). This work has received financial support from Ministerio de Ciencia, Innovaci\'on y Universidades/Agencia Estatal de Investigaci\'on, MICIU/AEI/10.13039/501100011033, Spain (PID2022-140510NB-I00, PCI2023-145952-2, RYC2019-027017-I, and Mar\'\i a de Maeztu grant CEX2023-001318-M); Xunta de Galicia, Spain (CIGUS Network of Research Centers and Consolidaci\'on 2021 GRC GI-2033 ED431C-2021/22 and 2022 ED431F-2022/15); and Feder Funds. We thank the NSF-funded White Mountain Research Station for their support. Computing resources were provided by the University of Chicago Research Computing Center and the Institute for Computational and Data Sciences at Penn State.

\section*{Full Author List: GRAND Collaboration}

\scriptsize
\noindent
J.~Álvarez-Muñiz$^{1}$, R.~Alves Batista$^{2, 3}$, A.~Benoit-Lévy$^{4}$, T.~Bister$^{5, 6}$, M.~Bohacova$^{7}$, M.~Bustamante$^{8}$, W.~Carvalho$^{9}$, Y.~Chen$^{10, 11}$, L.~Cheng$^{12}$, S.~Chiche$^{13}$
% ...
\\
\\
$^{1}$Departamento de Física de Particulas \& Instituto Galego de Física de Altas Enerxías, Universidad de Santiago de Compostela, 15782 Santiago de Compostela, Spain \\
$^{2}$Institut d'Astrophysique de Paris, CNRS  UMR 7095, Sorbonne Université, 98 bis bd Arago 75014, Paris, France \\
$^{3}$Sorbonne Université, Université Paris Diderot, Sorbonne Paris Cité, CNRS, Laboratoire de Physique 5 Nucléaire et de Hautes Energies (LPNHE), 6 4 place Jussieu, F-75252, Paris Cedex 5, France \\
$^{4}$Université Paris-Saclay, CEA, List,  F-91120 Palaiseau, France \\
$^{5}$Institute for Mathematics, Astrophysics and Particle Physics, Radboud Universiteit, Nijmegen, the Netherlands \\
$^{6}$Nikhef, National Institute for Subatomic Physics, Amsterdam, the Netherlands \\
$^{7}$Institute of Physics of the Czech Academy of Sciences, Na Slovance 1999/2, 182 00 Prague 8, Czechia \\
$^{8}$Niels Bohr International Academy, Niels Bohr Institute, University of Copenhagen, 2100 Copenhagen, Denmark \\
$^{9}$Faculty of Physics, University of Warsaw, Pasteura 5, 02-093 Warsaw, Poland \\
$^{10}$Key Laboratory of Dark Matter and Space Astronomy, Purple Mountain Observatory, Chinese Academy of Sciences, 210023 Nanjing, Jiangsu, China \\
$^{11}$School of Astronomy and Space Science, University of Science and Technology of China, 230026 Hefei Anhui, China \\
$^{12}$National Astronomical Observatories, Chinese Academy of Sciences, Beijing 100101, China \\
$^{13}$Inter-University Institute For High Energies (IIHE), Université libre de Bruxelles (ULB), Boulevard du Triomphe 2, 1050 Brussels, Belgium \\
% ...


\subsection*{Acknowledgments}

\noindent
The GRAND Collaboration is grateful to the local government of Dunhuag during site survey and deployment approval, to Tang Yu for his help on-site at the GRANDProto300 site, and to the Pierre Auger Collaboration, in particular, to the staff in Malarg\"ue, for the warm welcome and continuing support.
The GRAND Collaboration acknowledges the support from the following funding agencies and grants.
\textbf{Brazil}: Conselho Nacional de Desenvolvimento Cienti\'ifico e Tecnol\'ogico (CNPq); Funda\c{c}ão de Amparo \`a Pesquisa do Estado de Rio de Janeiro (FAPERJ); Coordena\c{c}ão Aperfei\c{c}oamento de Pessoal de N\'ivel Superior (CAPES).
\textbf{China}: National Natural Science Foundation (grant no.~12273114); NAOC, National SKA Program of China (grant no.~2020SKA0110200); Project for Young Scientists in Basic Research of Chinese Academy of Sciences (no.~YSBR-061); Program for Innovative Talents and Entrepreneurs in Jiangsu, and High-end Foreign Expert Introduction Program in China (no.~G2023061006L); China Scholarship Council (no.~202306010363); and special funding from Purple Mountain Observatory.
\textbf{Denmark}: Villum Fonden (project no.~29388).
\textbf{France}: ``Emergences'' Programme of Sorbonne Universit\'e; France-China Particle Physics Laboratory; Programme National des Hautes Energies of INSU; for IAP---Agence Nationale de la Recherche (``APACHE'' ANR-16-CE31-0001, ``NUTRIG'' ANR-21-CE31-0025, ANR-23-CPJ1-0103-01), CNRS Programme IEA Argentine (``ASTRONU'', 303475), CNRS Programme Blanc MITI (``GRAND'' 2023.1 268448), CNRS Programme AMORCE (``GRAND'' 258540); Fulbright-France Programme; IAP+LPNHE---Programme National des Hautes Energies of CNRS/INSU with INP and IN2P3, co-funded by CEA and CNES; IAP+LPNHE+KIT---NuTRIG project, Agence Nationale de la Recherche (ANR-21-CE31-0025); IAP+VUB: PHC TOURNESOL programme 48705Z. 
\textbf{Germany}: NuTRIG project, Deutsche Forschungsgemeinschaft (DFG, Projektnummer 490843803); Helmholtz—OCPC Postdoc-Program.
\textbf{Poland}: Polish National Agency for Academic Exchange within Polish Returns Program no.~PPN/PPO/2020/1/00024/U/00001,174; National Science Centre Poland for NCN OPUS grant no.~2022/45/B/ST2/0288.
\textbf{USA}: U.S. National Science Foundation under Grant No.~2418730.
Computer simulations were performed using computing resources at the CCIN2P3 Computing Centre (Lyon/Villeurbanne, France), partnership between CNRS/IN2P3 and CEA/DSM/Irfu, and computing resources supported by the Chinese Academy of Sciences.

\section*{Full Author List: Additional Authors}
\noindent Ingo Allekotte$^{1}$, Luciano Ferreyro$^{2}$, Matias Hampel$^{2}$, Federico Sanchez$^{2}$\\

\noindent $^{1}${Centro Atómico Bariloche and Instituto Balseiro (CNEA-UNCuyo-CONICET),San Carlos de Bariloche, Argentina}

\noindent $^{2}${Instituto de Tecnologías en Detección y Astropartículas (CNEA, CONICET, UNSAM), Buenos Aires, Argentina}

\end{document}